\pdfoutput=1
\def\dbcol{double column single spacing}
\ifdefined\dbcol
\documentclass[conference]{IEEEtran}
\IEEEoverridecommandlockouts
\else
\documentclass[a4paper]{article}
\fi

\usepackage{fancyhdr}

\ifdefined\COMSOC
  \usepackage[nocompress]{cite}
\else
  \usepackage{cite}
\fi
\ifdefined\ACM
	\usepackage{amsmath}
\else
	\usepackage[cmex10]{amsmath}
	\usepackage{amsthm}
	\usepackage[pdftex]{graphicx}
\fi

\ifdefined\META     \usepackage{stmaryrd}    \fi

\newcommand{\begproof}{\ifdefined\dbcol\begin{IEEEproof}\else\begin{proof}\fi}
\newcommand{\Endproof}{\ifdefined\dbcol\end{IEEEproof}\else\end{proof}\fi}

\usepackage[font=small]{subfig}

\usepackage{amssymb}
\usepackage{bbm}
\usepackage{verbatim}
\usepackage{color}
\usepackage[normalem]{ulem}

\usepackage{algorithmic}
\usepackage{algorithm}

\newtheorem{thm}{Theorem}
\newtheorem{lem}{Lemma}
\newtheorem{prop}{Proposition}
\newtheorem{coro}{Corollary}

\newcommand{\eref}[1]{Eq.~\eqref{#1}}
\newcommand{\fref}[1]{Fig.~\ref{#1}}
\newcommand{\tref}[1]{Table~\ref{#1}}
\newcommand{\sref}[1]{Section~\ref{#1}}
\newcommand{\thmref}[1]{Theorem~\ref{#1}}

\newcommand{\cref}[1]{Corollary~\ref{#1}}

\newcommand{\aref}[1]{Algorithm~\ref{#1}}

\newcommand{\vect}[1]{\boldsymbol{#1}}

\newcommand{\sumiN}{\sum_{i=1}^N}

\newcommand{\suml}{\sum_{l=1}^N}

\newcommand{\eqv}{\Leftrightarrow}
\newcommand{\means}{\Rightarrow}

\usepackage[table]{xcolor}

\title{Fairness and Social Welfare in \\
Incentivizing Participatory Sensing}
\author{\IEEEauthorblockN{Tie Luo$^*$\thanks{$^*$Tie Luo is now with the Networking Protocols Department, Institute for Infocomm Research (I2R), Singapore 138632. Email: luot@i2r.a-star.edu.sg.}}
\IEEEauthorblockA{Department of Electrical and Computer Engineering\\
National University of Singapore\\Singapore 117576\\
Email: luot@i2r.a-star.edu.sg}
\and
\IEEEauthorblockN{Chen-Khong Tham}
\IEEEauthorblockA{Department of Electrical and Computer Engineering\\
National University of Singapore\\Singapore 117576\\
Email: eletck@nus.edu.sg}
}

\vspace{-2cm}
\begin{document}
\maketitle

\pagestyle{fancy}
\thispagestyle{fancy}
\lhead{IEEE SECON 2012}
\cfoot{}
\rhead{\thepage}
\renewcommand{\headrulewidth}{0pt}

\begin{abstract}
Participatory sensing has emerged recently as a promising approach to large-scale data collection. However, without incentives for users to regularly contribute good quality data, this method is unlikely to be viable in the long run. In this paper, we link incentive to users' {\em demand} for consuming compelling services, as an approach complementary to conventional credit or reputation based approaches. With this demand-based principle, we design two incentive schemes, Incentive with Demand Fairness (IDF) and Iterative Tank Filling (ITF), for maximizing fairness and social welfare, respectively. Our study shows that the IDF scheme is max-min fair and can score close to 1 on the Jain's fairness index, while the ITF scheme maximizes social welfare and achieves a unique Nash equilibrium which is also Pareto and globally optimal. We adopted a game theoretic approach to derive the optimal service demands. Furthermore, to address practical considerations, we use a stochastic programming technique to handle uncertainty that is often encountered in real life situations.
\end{abstract}

\section{Introduction}\label{sec:intro}

With the vast penetration of smartphones with a variety of built-in sensors such as GPS, accelerometers and cameras, participatory sensing has emerged recently as a promising approach to large-scale data collection. Compared to the case of deployed sensors, participatory sensing removes the cost of installing and maintaining sensors and their inter-connection, while achieving much broader geographical coverage. The challenge of prolonging network lifetime in traditional sensor networks is no longer an issue in participatory sensing as the node battery is now taken care of by participating users themselves. Hence, participatory sensing is considered a promising new sensing paradigm and has attracted extensive attention and research efforts \cite{PEIR09mobisys,cmsense09sensys,lau11ccsa}.

However, the success of participatory sensing strongly relies on user participation to provide a sufficient and continuous influx of user contributions. So far, most participatory sensing studies, such as those mentioned above, focus on the sensing tasks per se, while the participating users are recruited on a voluntary or remunerated basis, which is not sustainable in the long run in real-life scenarios.

\begin{figure}[t]
\centering
\includegraphics[width=.5\textwidth]{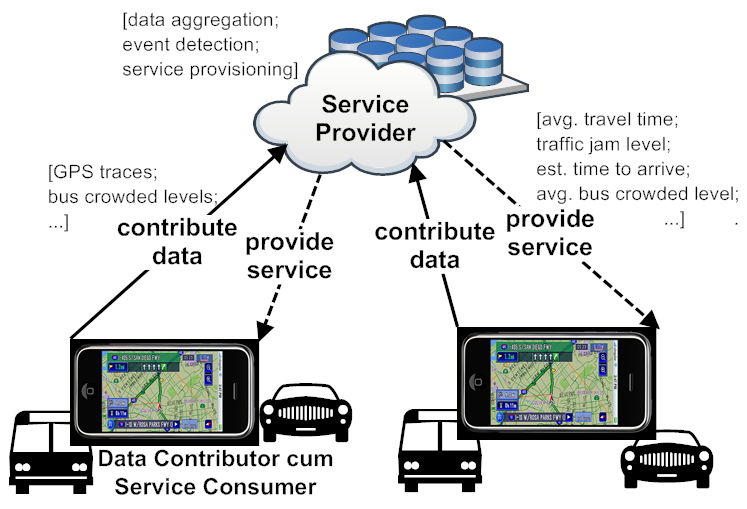}
\caption{An illustrative participatory sensing application: traffic monitoring.}
\label{fig:traffic}
\end{figure}

This brings forth the important issue of {\em incentive}, which acts as a driving force for participatory sensing. In this paper, we adopt a demand-based principle to design incentive schemes, by leveraging on the duality of a user's role: each user is a data {\em contributor} as well as a service {\em consumer}. 
We propose a participatory sensing framework consisting of data contributors and service consumers with a service provider that processes the contributed data and packages them into useful services that are consumed by service consumers. As an example, let us consider the case of traffic monitoring\cite{lau11ccsa} (see \fref{fig:traffic}). A smartphone user, when traveling on a bus or car, can act as a contributor of traffic data (e.g., GPS traces, bus crowd levels, etc.) to a service provider via a network connection (e.g., WiFi, GPRS, 3G, etc.). The service provider then aggregates and processes the data from all contributors and, thereby, provides a real-time traffic information service (e.g., browsing or querying on road jams, bus crowdedness, estimated time to reach destination etc.) for users to consume. Other applications include air pollution or noise level monitoring and flood or fire alerts, in which the required data are obtained from users' handset sensor readings or are entered by the users.

In all such cases, a user plays a dual role as a contributor as well as a consumer. Thus, we can exploit users' demand to consume useful and compelling services to design incentive schemes. In principle, a service provider will grant each user a {\em service quota}, which determines how much service he can consume, based on the user's consumption demand and his supply, i.e. contribution, level. The rationale is to provide, in addition to prior credit-based and reputation-based paradigms, another incentive design approach whereby making contributions is motivated by each user's intrinsic demand to consume services that, conversely, is based on all users' contributions.

When users' demand for services is of concern, there are two key questions to pose: How fairly will each individual user be satisfied? How well will all the users as a whole be satisfied? Accordingly, we consider two objectives in designing concrete incentive schemes: maximizing fairness and maximizing social welfare. For fairness, we design an incentive with demand fairness (IDF) scheme which is max-min fair and can score close to 1 on the Jain's fairness index. For social welfare, we design an iterative tank filling (ITF) scheme which achieves a unique Nash equilibrium that is simultaneously Pareto and global optimal, when a game-theoretic approach is taken. In addition, we use chance constrained programming, a stochastic programming technique, to handle uncertainty in real-life settings. Finally, we evaluate the performance of our schemes via simulations. Our results demonstrate the effectiveness of these schemes in meeting their respective objectives and confirm the theoretical results of our analysis.

\section{Related Work}

In the context of wireless ad hoc networks, incentive was studied as a means to stimulate each node to forward packets for other nodes, under the assumption that nodes are self-interested and try to conserve their own energy and transmission bandwidth. The approaches can be broadly classified into credit-based and reputation-based categories. For instance, Butty\'{a}n and Hubaux\cite{hubx03coop} proposed a credit-based mechanism using a virtual currency called nuglet: forwarding one packet for others will earn one nuglet while sending one own packet will consume one nuglet.
Marbach~\cite{marbach05}, on the other hand, added flexibility by allowing each node to freely decide on a forwarding price as well as sending rate in an adaptive manner. In reputation-based systems such as \cite{sori04}, each node's behavior is observed and evaluated by its neighbors and will further induce rewards or punishments based on the evaluation.

These approaches do not readily apply in our context of participatory sensing as users do not interact with each other directly but with a service provider (see \fref{fig:traffic}). Furthermore, our goal is not to stimulate cooperation among users but to attract more user contributions.

Recently, Park and van der Schaar\cite{park11infocom} introduced an intervention device that can take a variety of actions to influence users to cooperate and avoid {\em inefficiency}, under the assumption that the device can monitor a random access network, such as a CSMA network, perfectly. In our work, we focus on a different context, participatory sensing, and will show that our designed scheme achieves Pareto efficiency.

To the best of our knowledge, there are two studies that specifically address incentive for participatory sensing. Lee and Hoh\cite{hoh10pmc} proposed a dynamic pricing mechanism that allows users to sell their sensing data to a service provider. In order to keep the service provider's cost low while retaining an adequate number of participants, they proposed an auction mechanism to keep the bid price competitive while using ``virtual participation credits'' to retain participants. Our work does not use monetary incentive but leverages the dual role a user plays, thereby motivating users with their intrinsic demand for service. Furthermore, the techniques we use to tackle this problem are all different from those used by Lee and Hoh.
The other work was a pilot study conducted in UCLA by Reddy et al. \cite{reddy10micro} which looked into the effect of micro-payments in participatory sensing. The study found that monetary incentive is beneficial if combined with altruism and competitiveness, and participants were very concerned with fairness (which was left not addressed). In this paper, we not only use a different incentive scheme, but we also place great emphasis on and address the issue of fairness.

\section{System Model}\label{sec:model}

The system consists of $N$ users and a service provider. Each user is a data contributor as well as a service consumer. The service provider receives user contributed data and performs processing such as complex data mining or simple data fusion, at a server farm or computing cloud, and provides a value-added information service to the users. 

We consider the case where time is slotted. In each slot, a user $i\in[1..N]$ is characterized by a quadruple $\langle \psi_i, c_i, Q_i, q_i \rangle$, where $\psi_i$ is the user's contribution level in this slot,\footnote{There are various ways to evaluate a user's contribution level, which is application dependent. One such example is using the value of information (VoI) \cite{ipsn08community}.}
$c_i$ is the cost (e.g., mobile data charges and battery drainage) incurred by the user, $Q_i$ is the amount of service the user demands (in units of, e.g., hours) to consume in the {\em next} slot, which he declares anytime in the current slot, and $q_i$ ($q_i\le Q_i$) is the service quota that the service provider grants to the user (at the end of the current slot after calculating $\psi_i$), up to which he can actually use in the next slot.

In each slot, the service provider will provide a total amount $Q_{tot}$ of service quota, e.g., in units of person-hours, to all the users. This constant amount can either be fixed or vary from slot to slot; in the latter case, the service provider can leverage this to encourage more contribution by associating $Q_{tot}$ with the quality of service (QoS) of the system, denoted by $\Psi$, such that higher $\Psi$ will lead to higher $Q_{tot}$.\footnote{For example, let $Q_{tot} = \max\{1, \Psi/\Psi^*\}\times Q_{max}$ where $\Psi^*$ is the system-targetted QoS and $Q_{max}$ is the maximum service quota the system can provide (due to, e.g., system capacity and network bandwidth). As $\Psi$ is determined at the end of each slot, $Q_{tot}$ is also determined then and will be consumed in the next slot.} The QoS $\Psi$ is determined by the aggregate amount of all users' contribution. Since the expression of $\Psi$ is application dependent, so for the sake of generality, our subsequent analysis will not be coupled with any specific expression of $\Psi$. Rather, we only assume that QoS is positively correlated to the amount of users' contributions and that a single user does not noticeably affect QoS. In other words, $\Psi$ can be viewed as constant with respect to an individual $\psi_i$ provided that the user population is sufficiently large.

Therefore, the problem is to assign an amount $Q_{tot}$ of service quota to $N$ users according to their characterizing quadruple, with the objective of maximizing fairness (\sref{sec:fair}) or social welfare (\sref{sec:welfare}).

\ifdefined\JNL\blueul{Formulate these two objectives, maybe in respective sections.}\fi

\section{Incentive with Fairness}\label{sec:fair}

From an individual user's perspective, one would expect a ``fair'' rewarding scheme when consuming the service. In accordance with the demand-based principle, the fairness here is defined as that a user $i$'s received service quota $q_i$ is commensurate with both his contribution level $\psi_i$ and his demand $Q_i$.\footnote{Contribution cost $c_i$ will be considered in \sref{sec:welfare}.} This section designs an incentive scheme called Incentive with Demand Fairness (IDF) to achieve this objective. 

Let us first consider a simpler incentive scheme which grants $q_i$ based on $\psi_i$ only. To do this, we gradually increase $q_i$ for each user $i$ at the differentiated rate of $\psi_i/\suml\psi_l$. Once a user's demand $Q_i$ is reached, we exclude this user and proceed with the rest of the users in the same way, but with an updated $Q_{tot}$. This process can be mathematically described, by re-indexing the users by $j=1,...,N$ in ascending order of $Q_i/\psi_i$ (which is the order in which the users will be fully satisfied), as:
\begin{align}\label{eq:absq}
q_j = \min \{ Q_j, \frac{\psi_j}{\sum_{k=j}^N\psi_k}\times (Q_{tot} - \sum_{k=1}^{j-1} q_k)\}.
\end{align}

Next, we consider the scheme by also taking demand into account. Naturally, it is fair to grant $q_i$ such that $q_i/Q_i$ is proportional to $\psi_i$ when neither of the limits, $Q_i$ and $Q_{tot}$, is reached. One way to do this is, by mimicking the case above, to increase $q_i/Q_i$ at the rate of $\psi_i/\suml\psi_l$ until 1 is reached. However, this does not readily lead to a mathematical or algorithmic abstraction because, unlike in \eref{eq:absq} where $\sumiN q_i$ is capped by $Q_{tot}$, the upper bound to $\sumiN q_i/Q_i$ is not clear when $\sumiN Q_i>Q_{tot}$. Therefore, instead, we increase each $q_i$ at the rate of $Q_i\psi_i/\suml Q_l\psi_l$ until reaching $Q_i$, whereby the user with the largest $\psi_i$ will obtain the maximal $q_i/Q_i$ first, which fulfills the objective. Thus, the scheme can be formulated below, by sorting the users in descending order of $\psi_i$ and re-indexing them by $j=1,...,N$:
\begin{align}\label{eq:relq}
q_j = \min \{ Q_j, \frac{Q_j\psi_j}{\sum_{k=j}^N Q_k\psi_k} (Q_{tot}-\sum_{k=1}^{j-1}q_k)\}.
\end{align}
This is the IDF scheme which is algorithmically presented as \aref{alg:idf}.
\begin{algorithm}[ht]
\caption{Incentive with Demand Fairness (IDF)}\label{alg:idf}
\begin{algorithmic}[1]
\REQUIRE $N,Q_{tot},\vec Q=\{Q_i\},\vec\psi=\{\psi_i\}$
\ENSURE $\vec q=\{q_i\}$
\IF{$\sumiN Q_i\le Q_{tot}$}
\RETURN $\vec q\gets\vec Q$
\ENDIF
\STATE create $\vec I=\{I_j\}_{j=1}^N$, $I_j\in[1,N]$, such that $\psi_{I_1},\psi_{I_2},...\psi_{I_N}$ are in descending order
\FOR{$j=1\to N$}
	\STATE $q_{I_j}=\displaystyle Q_{tot} \frac{Q_{I_j}\psi_{I_j}}{\sum_{k=j}^N Q_{I_k}\psi_{I_k}}$
	\IF{$q_{I_j}>Q_{I_j}$}
	\STATE $q_{I_j}\gets Q_{I_j}$
	\ENDIF
	\STATE $Q_{tot}-=q_{I_j}$
\ENDFOR
\end{algorithmic}
\end{algorithm}

To analyze the properties of IDF, we consider two important and well-established fairness measures, Jain's fairness index\cite{jain98} and max-min fairness. Jain's fairness index is defined as
\[ J=\frac{(\sumiN x_i)^2}{N \sumiN x_i^2}\]
where, in our context, $x_i\triangleq q_i/q_i^*$ in which $q_i^*$ is the optimal (i.e., fairest) service quota to be granted to user $i$. The maximum of Jain's fairness index is 1, achieved when $x_i=x_j, \forall i,j$. In line with our objective of fairness, $q_i^*=Q_i\psi_i$ (ignoring a constant coefficient which does not affect the result). Thus,
\begin{align}\label{eq:jrsq}
J = \frac{(\sumiN\frac{q_i}{Q_i\psi_i})^2}{N\sumiN(\frac{q_i}{Q_i\psi_i})^2}.
\end{align}

Without loss of generality, suppose there are $k$ ($0\le k\le N$) users who are fully satisfied, and all the users are sorted as in \eqref{eq:relq} and indexed by $j$. It is fairly straightforward to show that
\begin{align}
q_j=\begin{cases}
Q_j, &j=1,...,k\nonumber\\
h\times Q_j\psi_j, &j=k+1,...,N\nonumber
\end{cases}
\end{align}
where
\[h= \frac{Q_{tot} - \sum_{j=1}^k Q_j}{\sum_{j=k+1}^N Q_j\psi_j}\]
and $k$ is determined by $1/\psi_k\le h<1/\psi_{k+1}$. Hence,
\begin{align}\label{eq:jidf}
J = \frac{[\sum_{j=1}^k \psi_j^{-1} + (N-k)h]^2}{N [\sum_{j=1}^k \psi_j^{-2} + (N-k) h^2]}.
\end{align}

We will evaluate \eref{eq:jidf} for IDF and several other schemes in \sref{sec:simu} through simulations. Here, we give two special simple cases that can be theoretically solved:
\begin{itemize}
\item $k=0\means J=1$: in this case, the maximum fairness is achieved, and all users are equally satisfied. The expression for $h$ is $h = q_i/(Q_i\psi_i) = Q_{tot}/\suml Q_l\psi_l$.
\item $k=N\means J = (\sumiN\psi_i^{-1})^2/(N \sumiN\psi_i^{-2})$: in this case, all users are {\em fully} satisfied, and $J=1$ if all the users contribute equally.
\end{itemize}

The result for the other fairness measure, max-min fairness, is given below.
\begin{prop}
The IDF scheme achieves weighted max-min fairness. That is, increasing user $i$'s demand-normalized service quota, $q_i/Q_i$, weighted by $1/\psi_i$, viz. $q_i/(Q_i\psi_i)$, must be at the cost of decreasing some other user $j$'s $q_j/(Q_j\psi_j)$, where $q_j/(Q_j\psi_j)<q_i/(Q_i\psi_i)$. 
\end{prop}

\section{Incentive with Social Welfare Maximization}\label{sec:welfare}

From a system perspective, we consider the objective that the service provider aims to maximize social welfare, where social welfare is defined as the aggregate user utility with respect to the service provided by the system. In the meantime, the system shall also incentivize users to contribute at higher levels. Therefore, the objective is formulated as maximizing $S\triangleq\sumiN\psi_i u_i$, the aggregate contribution-weighted user utility, where $u_i$ is user $i$'s utility. The structure of this objective function implies that priority will be given to users with larger $\psi_i$.

The utility $u_i$ can be defined as one of two possible forms:
\begin{align}\label{eq:utdef}
(a)\;u_i=U(\Psi \frac{q_i}{c_i Q_i}),\hspace{1cm}(b)\;u_i=U(\Psi \frac{q_i}{Q_i})/c_i.
\end{align}
In (a), $q_i/(c_i Q_i)$ is a user's demand-normalized service quota ($q_i/Q_i$) evaluated against cost $c_i$. $\Psi$ is the system QoS, as described earlier. $U(x):\mathbb R^+\to\mathbb R$ is a utility function monotonically increasing and strictly concave in $x$, which reflects the {\em elasticity} of user satisfaction as is common in the literature.
In this paper, we consider the form of $U(x)=\log(1+x), x\ge 0$ \cite{marbach05,ma06p2p,shenker95jsac}, and thus the problem is formulated below as a nonlinear programming problem:
\begin{align}
\text{maximize: (a) }S&=\sumiN\psi_i\log(1+\Psi\frac{q_i}{c_i Q_i}),\text{ or}\label{eq:swa}\\
\text{maximize: (b) }S&=\sumiN\psi_i\log(1+\Psi\frac{q_i}{Q_i})/c_i,\label{eq:swb}\\
\text{s.t.\ }&q_i\in[0,Q_i],\ \forall i=1,...,N\\
&\sumiN q_i\le Q_{tot}.\label{c:tot}
\end{align}

\begin{figure}[ht]
\centering
\includegraphics[trim=3mm 4mm 2mm 3mm,clip,width=0.35\textwidth]{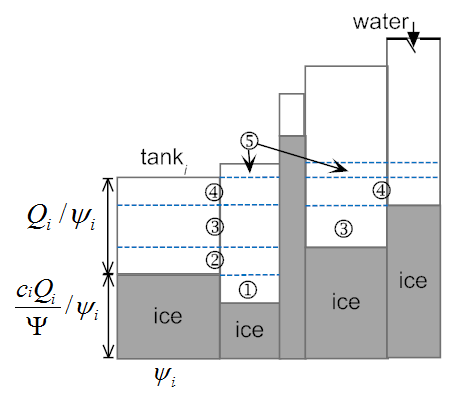}
\caption{Filling iced tanks. The tanks are sorted in ascending order of height, but the ice (gray area) levels are not necessarily in order. Numbers \textcircled{1},\textcircled{2},\textcircled{3},... denote the order of empty spaces to be filled, which also correspond to the iteration number of the ITF algorithm.}
\label{fig:tank}
\end{figure}

In this section, we design a scheme to solve problem (a) while leaving problem (b) to \cite{ps11ext} for interested readers, since both problems follow the same line of reasoning and (b) turns out to be simpler than (a). Now, let us consider \eref{eq:swa}. In order to maximize $S$, the solution should give priority to users with larger {\em marginal} weighted utility, i.e., larger $\psi_i \Psi/(c_i Q_i+q_i \Psi)$ ($q_i$ being the optimizing variables), or equivalently, smaller $(\frac{c_i Q_i}{\Psi}+q_i)/\psi_i$. With this point of view, we convert the original NLP problem into a problem of ``filling iced tanks'' depicted in \fref{fig:tank}. Each user $i$ is represented by a tank with bottom area $\psi_i$, and the tank has been preoccupied by frozen ice of volume $c_i Q_i/\Psi$ (and hence of height $\frac{c_i Q_i}{\Psi}/\psi_i$). Tank $i$ is left with an empty space of volume $Q_i$ (and hence of height $Q_i/\psi_i$) to be filled with water. All the tanks are placed back to back as if they are virtually connected without internal separators. Consequently, the empty space will be filled consecutively in the order of \textcircled{1},\textcircled{2},\textcircled{3},... shown in \fref{fig:tank}. To solve the problem,\footnote{This iced-tank filling problem is different from the water filling (WF) problem in convex optimization \cite{boyd04}
or wireless communications \cite{tse05wc}
in that (i) these tanks can have different water levels during and after filling, because each tank comes with a closed ``lid'' due to the constraint $q_i\le Q_i$, whereas WF fills a single and open vessel with one sweeping water level, (ii) WF will fully allocate the total resource (power) which however is not the case in ITF.} we design an {\em iterative tank filling} (ITF) algorithm which iteratively fills the space in the depicted order until all the tanks are fully filled or the total volume of water, $Q_{tot}$, is used up. The pseudo-code is given in Algorithm~\ref{alg:itf}.
\begin{algorithm}[ht]
\caption{Iterative Tank Filling (ITF)}\label{alg:itf}
\begin{algorithmic}[1]
\REQUIRE $N,Q_{tot},\Psi,\vec Q=\{Q_i\},\vec\psi=\{\psi_i\},\vec c=\{c_i\}$
\ENSURE $\vec q=\{q_i\}$
\IF{$\sumiN Q_i\le Q_{tot}$}
\RETURN $\vec q\gets\vec Q$
\ENDIF
\STATE $\vec q\gets \vec 0$; $\vect{ice}\gets\{ice_i=c_i Q_i/(\psi_i\Psi)\}$;\\
$\vect{tank}\gets\{tank_i=c_i Q_i/(\psi_i\Psi)+Q_i/\psi_i\}$\label{alg:init}
\WHILE{$Q_{tot} >0$}
\STATE ------------ Find space to fill in this iteration ------------
\STATE $bot\gets\min_i\{ice_i\};\;botind\gets\arg\min_i \{ice_i\}$\label{alg:bot}
\STATE $w\gets\sum_{i\in botind} \psi_i$ //bottom area
\STATE $cap_1\gets\min_{i\notin botind}\{ice_i\}$
\STATE $cap_2\gets\min_i\{tank_i\}$\label{alg:cap2}
\STATE $h\gets\min\{cap_1, cap_2 \}-bot$ //height
\STATE ----- {Fill the space which may span multiple tanks} -----
\IF{$w\cdot h<Q_{tot}$}
\STATE $Q_{tot} -= w\cdot h$
\ELSE [the last iteration of filling]
\STATE $h \gets Q_{tot}/w$ //readjust height
\STATE $Q_{tot}\gets 0$ \label{alg:Q0}
\ENDIF
\FORALL{$i\in botind$}
\STATE $ice_i+=h;\; q_i+=h\cdot \psi_i$ \label{alg:fill}
\ENDFOR
\STATE --------------------- {Remove full tanks} -----------------------
\IF {$cap_2\le cap_1$ or $cap_1=\infty$}
\FORALL{$k\in\{i|tank_i=cap_2\}$}
\STATE $tank_k\gets\infty$; $ice_k\gets\infty$ \label{alg:remfull}
\ENDFOR
\ENDIF
\ENDWHILE
\end{algorithmic}
\end{algorithm}

\begin{prop}
The computational complexity of ITF is $O(N^2)$.
\end{prop}
\begin{IEEEproof}
In the worst case, $ice_i$ and $tank_i$ are all different (i.e., 2$N$ distinct numbers), and hence each iteration will increase the highest water level to only one of these 2$N$ numbers. Therefore, The main loop will execute at most $2N-1$ times. Inside the main loop, lines \ref{alg:bot}-\ref{alg:cap2}, \ref{alg:fill}, and \ref{alg:remfull} each has a complexity of $O(N)$. The proposition is thus proven.
\end{IEEEproof}

\begin{thm}\label{thm:highut}
Service provisioning via ITF ensures that, for any $i,j$, if $\displaystyle\frac{\psi_i}{c_i Q_i}\ge\frac{\psi_j}{c_j Q_j}\land c_i\le c_j$, then $u_i\ge u_j$.
\end{thm}
\begin{figure}[ht]
\subfloat[Case 1.]{\label{fig:case1}
\includegraphics[trim=3mm 4mm 3mm 2mm,clip,width=0.48\linewidth]{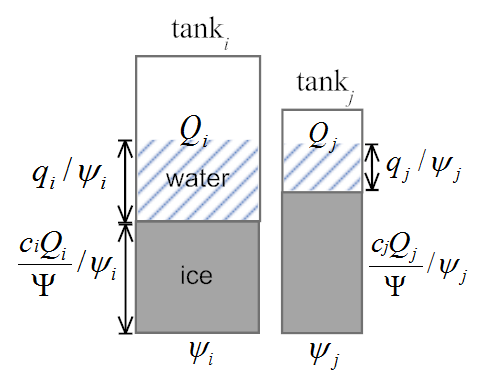}}\hfil
\subfloat[Case 2.]{\label{fig:case2}
\includegraphics[trim=3mm 4mm 3mm 2mm,clip,width=0.48\linewidth]{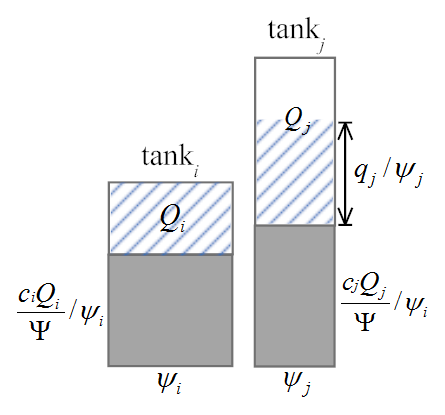}}
\caption{Proof of Theorem~\ref{thm:highut}.}
\end{figure}
\begin{IEEEproof}
Consider two cases of the output $\vec q$:
\begin{enumerate}
\item $q_i/\psi_i\ge q_j/\psi_j$ (Fig. \ref{fig:case1}). Multiplying this with $\frac{\psi_i}{c_i Q_i}\ge\frac{\psi_j}{c_j Q_j}$ gets $\frac{q_i}{c_i Q_i}\ge\frac{q_j}{c_j Q_j}$. Hence $U(\Psi\frac{q_i}{c_i Q_i})\ge U(\Psi\frac{q_j}{c_j Q_j})$ (for any non-decreasing function $U(\cdot)$), i.e., $u_i\ge u_j$.
\item $q_i/\psi_i<q_j/\psi_j$. Since $\frac{\psi_i}{c_i Q_i}\ge\frac{\psi_j}{c_j Q_j} \eqv \frac{c_i Q_i}{\Psi}/\psi_i\le\frac{c_j Q_j}{\Psi}/\psi_j$, meaning that the reciprocal of marginal weighted utility or the ice level of $i$ is lower than that of $j$, priority will be given to $i$ (ITF will start filling tank $i$ earlier than $j$). However, since the outcome is $q_i/\psi_i<q_j/\psi_j$, it implies that tank $i$ must have been fully filled and the height of original empty space $Q_i/\psi_i<Q_j/\psi_j$, as shown in \fref{fig:case2}. As $c_i\le c_j$, we have $1/c_i=\frac{q_i}{c_i Q_i}\ge1/c_j\ge\frac{q_j}{c_j Q_j}\Rightarrow\log(1+\frac{\Psi}{c_i})\ge \log(1+\Psi\frac{q_j}{c_j Q_j})\Rightarrow u_i\ge u_j$.
\end{enumerate}
\end{IEEEproof}
\begin{coro}\label{cor:highut}
Service provisioning via ITF ensures that, for any $i,j$, if $\displaystyle\frac{\psi_i}{Q_i}\ge\frac{\psi_j}{Q_j}\land c_i\le c_j$, then $u_i\ge u_j$.
\end{coro}
\cref{cor:highut}, as a relaxed form of \thmref{thm:highut}, shows that a user who makes higher contribution with respect to his demand and incurs lower cost, will be guaranteed higher utility.

For completeness, we briefly explain how the service provider calculates users' costs, $\vec c$. The major component of $c_i$ is the mobile data charge incurred by the user when making a contribution. This can be calculated using the user's mobile data plan obtained from the user's registration information, the time of the contribution and the amount of data contributed, which can be easily measured at the server. The other minor component is the user's battery drainage, which can be gauged from the user's phone model and amount of data contributed. The service provider can then feed back the calculated cost $c_i$ to the corresponding user.

Expanding on our proposed demand-based approach, we now address two other major issues in the following sub-sections.

\subsection{Optimal Service Demands}\label{sec:game}

One issue is to derive the optimal service demands, $Q_i$, that users declare. This is of interest because of the following. As each user's service quota is capped by his {\em declared} demand, a user may be tempted to declare a higher demand in order to, possibly, get a larger share of service quota. On the other hand, as an incentive scheme should be {\em transparent} to users, a user can realize from ITF that declaring a higher demand can, conversely, put him into a disadvantageous situation in which he will be classified as a ``hard-to-satisfy'' user and given a lower priority to receive service. Therefore, there should exist an {\em optimal} service demand for each user.

There are two ways to define the optimality: (i) global optimality---the objective function \eqref{eq:swa} achieves the maximum over the entire domain of optimization variables; (ii) Pareto optimality---no user's utility can be improved without making some other user's utility worse off. These two kinds of optimality are {\em not} achieved simultaneously in general.

In addition, it is also desirable to make the optimal point ``stable'': any user should not have incentive to deviate from his optimal demand unilaterally, i.e., if other users stick to their optimal demands.

Under these circumstances, a game-theoretic approach is appropriate. This sub-section derives the solution and shows that it achieves all of the aforementioned properties: global maximal, Pareto optimal, and Nash Equilibrium.

We model the participatory sensing problem as a non-cooperative game \cite{game07ras}. The game players are the $N$ users. Each player's strategy is to decide how much demand, i.e., $Q_i$, to declare and his strategy space is $\mathbb R^+$. Each user's payoff is his utility (by receiving the service quota granted by the service provider). The game rule, prescribed by ITF, maps a $N$-tuple of user strategies $\vec Q=\{Q_i\}\in(\mathbb R^+)^N$ to a $N$-tuple of user payoffs $\vec u=\{u_i\}\in(\mathbb R^+)^N$, by determining the service quota $\vec q$.

A common game-theoretic approach is to find a strategy profile, prove it to be a Nash Equilibrium (NE), and subsequently prove uniqueness if possible. We take a different approach: we first find the necessary and sufficient conditions for a NE, and then derive the NE and prove uniqueness in one go.
\begin{lem}\label{thm:necond}
The necessary and sufficient conditions that a Nash Equilibrium of the above-defined game satisfies are
\begin{align}\label{eq:necond}
\begin{cases}{\rm C1:\;\;}\sumiN Q_i = Q_{tot}\\
{\rm C2:\;\;}h_i=h_j,\;\forall i,j=1,...,N\end{cases}
\end{align}
where $h_i=Q_i/\psi_i + c_i Q_i/(\Psi\psi_i)$.
\end{lem}
\begin{IEEEproof} Necessity: Prove by contradiction as follows.\\
Condition C1: Suppose, instead, $\sumiN Q_i<Q_{tot}$, then $q_i=Q_i, \forall i$. Obviously, one can increase his $Q_i$ to $Q_i'>Q_i$ and be granted $q_i'>q_i$ provided that other users do not change strategy. If, otherwise, $\sumiN Q_i>Q_{tot}$, there is at least one tank that is not fully filled. Let $k$ be one such tank and $b_k=c_k Q_k/(\Psi\psi_k)$ denote its ice level. Let $h_i$ be tank $i$'s {\em ice+water} level and $h_{max}=\max_i\{h_i|q_i>0\}$. 
In one case that $b_k\ge h_{max}$ (i.e., $q_k=0$), clearly user $k$ can decrease demand $Q_k$ such that $b_k$ drops to $b_k'<h_{max}$, and be granted $q_k'>0$. In the other case that $b_k<h_{max}$ (i.e., $0<q_k<Q_k$), $k$ can decrease $Q_k$ to $Q_k'$ such that $q_k<Q_k'<Q_k$ and, accordingly, ice level $b_k$ drops to $b_k'=c_k Q_k'/(\Psi\psi_k)$,
and be granted: (i) $q_k'>q_k$ if $\exists j: q_k/\psi_k+b_k'<h_j\le h_{max}$, where the left hand side ($q_k/\psi_k+b_k'$) is $k$'s ice+water level as if his water level remains unchanged, or (ii) $q_k'=q_k$ otherwise (such $j$ does not exist; i.e., $k$ is the only partially-filled tank ($0<q_k<Q_k$) and the rest of the tanks are either fully filled (with their $h_j\le q_k/\psi_k+b_k'$) or empty (with their $h_j>h_{max}$)).  
In summary, user $k$ will {\em have} an incentive to deviate from his strategy if $\sumiN Q_i\ne Q_{tot}$. Therefore, $\sumiN Q_i = Q_{tot}$ must hold.

Condition C2: Suppose $\exists i,j:h_i\ne h_j$, and WLOG, $h_i<h_j$. Since $\sumiN Q_i = Q_{tot}$, all the users are fully satisfied. Recall that $h_i$ and $h_j$ are the ice+water level of users $i$ and $j$, respectively. If user $i$ increases his demand (slightly) to $Q_i'$ such that $h_i'=Q_i'/\psi_i + c_i Q_i'/(\Psi\psi_i)<h_j$ still holds, then, according to the ITF game rule, user $i$ will be granted $q_i'>q_i=Q_i$ where the additional quota essentially comes from user $j$ (and others, if any). This means that user $i$ will have an incentive to change his strategy unilaterally. Hence, Condition 2 must also hold.

Sufficiency:\\
If both C1 and C2 are satisfied, all the tanks have the same height and are fully filled. Suppose any user, say $i$, changes his strategy such that: (1) $Q_i'>Q_i$, then $i$'s ice level will increase, which actually lowers $i$'s priority to receive service. On the other hand, all the other tanks are fully filled. Hence, tank $i$ will continue to have the same volume, $Q_i$, of water (though its ice+water level will be above the other tanks) with an empty space of $Q_i'-Q_i$ left in the tank; (2) $Q_i'<Q_i$, then obviously he will receive a {\em lower} quota of $q_i'=Q_i'$. In summary, user $i$ will either be indifferent (in case 1) or unwilling (in case 2) to switch his strategy. Hence, a strategy profile satisfying both C1 and C2 is a NE.
\end{IEEEproof}
\begin{thm}\label{thm:ne}
The optimal strategy profile $\vect Q^*=\{Q_i^*\}$ where
\begin{align}\label{eq:qne}
Q_i^* = \frac{\psi_i/(\Psi+c_i)}{\suml\frac{\psi_l}{\Psi+c_l}}Q_{tot}
\end{align}
is a unique Pareto-efficient Nash equilibrium, and achieves the global optimum.
\end{thm}
\begin{IEEEproof}
It can be shown that the equation system \eqref{eq:necond} can be translated into a $N\times N$ homogeneous system of linear equations whose determinant is non-zero. Hence, this system has a unique solution which is then obtained to be \eref{eq:qne}.

The Pareto efficiency follows from C1 of Lemma~\ref{thm:necond}.

Under the NE strategy \eqref{eq:qne}, $q_i=Q_i$ and each user receives the maximum utility $u_i^{max}=\log(1+\Psi/c_i)$ for given $\vec \psi$ and $\vec c$. This achieves the global maximum of \eqref{eq:swa} {\em term-wise}, which is a sufficient condition for \eqref{eq:swa} to achieve its global maximum:
\begin{align}\label{eq:objane}
S_a^{max}=\sumiN \psi_i\log(1+\Psi/c_i).
\end{align}
\end{IEEEproof}

In practice, the service provider shall announce each user's contribution level $\psi_i$ and cost $c_i$, as well as the total quota $Q_{tot}$ and system QoS $\Psi$, for each user to calculate his optimal $Q_i^*$.

\subsection{Uncertainties in Service Demands}\label{sec:ccp}
The other issue is to handle uncertainty which is commonly encountered in reality. As demands are essentially future demands, or specifically, $Q_i$ is the amount of service a user plans to consume in the next slot, a user needs to {\em estimate} his actual demand. Denote the (unknown) actual demand in the next slot by $\tilde Q_i$, which is a random variable, and the estimated demand by $\hat Q_i$. Thus, the previously discussed $Q_i$ is actually $\hat Q_i$, and $\hat Q_i$ is the ``expected value'' of $\tilde Q_i$. To reformulate the problem by taking the actual demand $\tilde Q_i$ into account, it is improper to replace the original constraint $q_i\le Q_i$ with $q_i\le\tilde Q_i$ which essentially leads to $q_i\le\inf\{\tilde Q_i\}=0$. It is also improper to replace $q_i\le Q_i$ with $q_i\le\hat Q_i$ which will be elaborated in \sref{sec:evm}. The proper way is to introduce probabilistic constraints, such as
\begin{align}\label{c:icc}
\Pr(q_i\le \tilde Q_i)\ge 1-\alpha_i,\;\forall i=1,...,N
\end{align}
where $\alpha_i$'s are prescribed probabilities. Each of these $N$ constraints means that $q_i$ is capped by (all the realizations of) $\tilde Q_i$ in $1-\alpha_i$ of the time, or alternatively, $q_i$ has a chance of $\alpha_i$ to exceed $\tilde Q_i$. 

\begin{figure*}[tb]
\subfloat[EVM. 45 cases of over-provisioning.]{\label{fig:expvalue}\includegraphics[width=0.32\textwidth]{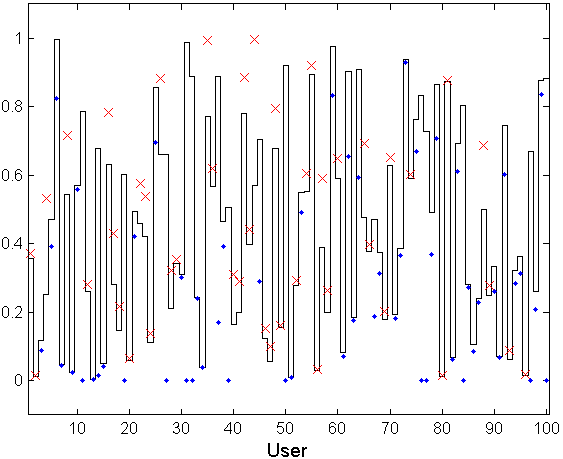}}\hfil
\subfloat[EVM. Zoom-in view of users 1--10.]{\label{fig:expvalue10}\includegraphics[width=0.32\textwidth]{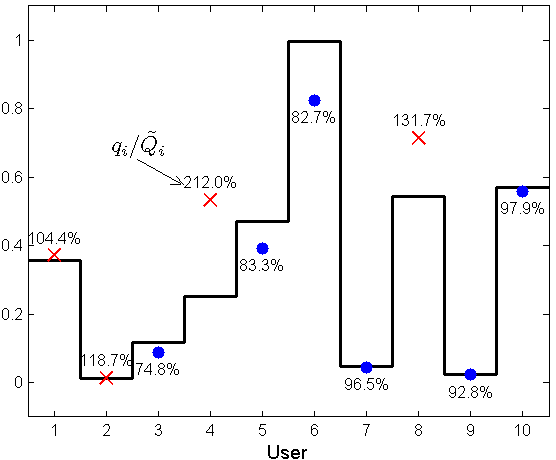}}\hfil
\subfloat[CCP method. 4 cases of over-provisioning.]{\label{fig:ccp}\includegraphics[width=0.32\textwidth]{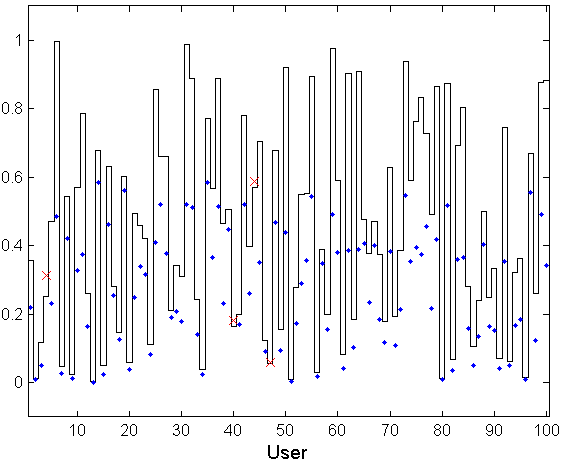}}
\caption{Solving the chance constrained problem~\eqref{eq:objaccp} using EVM and CCP, respectively. $N$=100 users. The staircase line represents a realization of all the $\tilde Q_i$'s. Blue dots are cases where $q_i\le\tilde Q_i$ and red crosses are $q_i>\tilde Q_i$ (over-provisioning). The vertical axis stands for $\tilde Q_i$ and $q_i$.}
\end{figure*}

The inequality \eqref{c:icc} imposes {\em individual chance constraints} on the objective function, where ``individual'' relates to the fact that each stochastic constraint $q_i<\tilde Q_i$ is transformed into a chance constraint individually.
A variant is called {\em joint chance constraints} which, however, does not capture our problem as well as \eqref{c:icc}. Hence, we leave the discussion to \cite{ps11ext} for interested readers.

The original problem can be then reformulated as:
\begin{equation}\label{eq:objaccp}
\begin{aligned}
\text{maximize }S=&\sumiN\psi_i\log(1+\Psi \frac{q_i}{c_i \hat Q_i}),\\
\text{s.t.\;\;\;}&\Pr(q_i\le \tilde Q_i)\ge 1-\alpha_i,\;\forall i=1,...,N,\\
&q_i\ge 0,\;\;\;\forall i=1,...,N,\\
&\sumiN q_i\le Q_{tot}.
\end{aligned}
\end{equation}
The objective function uses $\hat Q_i$ instead of $\tilde Q_i$ (which is a random variable), because a user's utility is determined when he is granted $q_i$ based on his declaration $\hat Q_i$, instead of $\tilde Q_i$. We also note that, with the introduction of the chance constraints \eqref{c:icc}, a service provider has the option of giving another incentive by associating $\alpha_i$ with user contribution, e.g., letting $\alpha_i=\alpha_0\psi_i/\suml \psi_l$ where $\alpha_0$ is a scaling factor.

Assuming $\tilde Q_i\sim\mathcal N(\hat Q_i,\sigma_i)$, we set out to solve the stochastic programming problem \eqref{eq:objaccp}.

\subsubsection{Expected-Value Method}\label{sec:evm}
One may be of the opinion that the new formulation \eqref{eq:objaccp} is not much different from replacing constraint \eqref{c:icc} by $q_i\le\hat Q_i$, which straightforwardly converts \eqref{eq:objaccp} into the original (deterministic) optimization problem whose solution is already given by ITF. This is called an expected-value method (EVM) which uses $\mathbb E(\tilde Q_i)=\hat Q_i$ to simplify the constraints. 

To examine whether EVM is suitable for our particular problem, we conducted a simulation study for an hourly-slotted system with $N$=100 users. For ease of description, denote by $\mathcal U(a,b)$ the uniform distribution in interval ($a,b$), and by $\mathcal N_{tr}(\mu,\sigma,a,b)$ the truncated normal distribution with mean $\mu$ and standard deviation $\sigma$ and bounded in the range of [$a,b$]. In the simulation setup, $\hat Q_i=\mathcal U(0,1)$ (hour), $\tilde Q_i=\mathcal N_{tr}(\hat Q_i,0.25\hat Q_i,0,1)$, contribution $\psi_i=\mathcal U(0,1)$ (kb), and cost $c_i=\mathcal N_{tr}(\psi_i,\psi_i,0,3)$ (the expected cost is one dollar per kb of contribution). The total service quota $Q_{tot}=0.75\sumiN\hat Q_i$
and QoS $\Psi=10^{-3}\sumiN\psi_i$ (Mb).

According to EVM, we simply replace the original $Q_i$ with $\hat Q_i$ to run the ITF algorithm. The results are shown in \fref{fig:expvalue}, where 45 out of 100 users were found to have exceeded their actual demands $\tilde Q_i$, which results in significant resource wastage since a user would not consume more than his actual demand. Zooming on users 1--10, we see in \fref{fig:expvalue10} that $q_i/\tilde Q_i$ can be as high as 212\%. Therefore, these observations convey that EVM can cause significant {\em service over-provisioning} and, perhaps more importantly, the service provider has {\em no control} over such over-provisioning. The consequence is that users will {\em lose incentive} because any user is likely to be granted a large share of service {\em without} commensurate contribution, which works strongly against any incentive scheme.

\subsubsection{Chance Constrained Programming}
Now that we have seen that EVM is not suitable for our particular problem with uncertainty, we use the {\em chance constrained programming} (CCP)\cite{ap95sp} approach to tackle it.

For constraint \eqref{c:icc}, denote the CDF of $\tilde Q_i$ by $F_i(\cdot)$,
\[\Pr(q_i\le \tilde Q_i)\ge 1-\alpha_i \eqv F_i(q_i)\le \alpha_i.\]
Denote by $\gamma_i(p)$ the quantile function of $Q_i$, defined as
\[\gamma_i(p)\triangleq \inf\{\tau|F_i(\tau)\ge p\}.\]
As $F_i(\cdot)$ is monotonically increasing, it follows that
\begin{align}\label{eq:zalpha}
F_i(q_i)\le\alpha_i \Leftrightarrow q_i\le\gamma_i(\alpha_i).
\end{align}
In order to solve for $\gamma_i(\alpha_i)$, we use the {\em probit function} which is the quantile function for the {\em standard} normal distribution and can be computed via easy numerical computation or simple table look-up.
Denote by $z_{\alpha}$ the $\alpha$-quantile of the standard normal distribution. Since $\tilde Q_i\sim\mathcal N(\hat Q_i,\sigma_i)$, we have $(\tilde Q_i-\hat Q_i)/\sigma_i\sim\mathcal N(0,1)$, and hence \eqref{eq:zalpha} can be transformed into
\begin{align}\label{eq:iccdone}
\frac{q_i-\hat Q_i}{\sigma_i}\le z_{\alpha_i}\eqv
q_i\le \hat Q_i+\sigma_i z_{\alpha_i},
\end{align}
where $z_{\alpha_i}$ can be obtained via numerical computation or standard table lookup, e.g., $z_{0.05}=-1.65, z_{0.025}=-1.96$.

Thus, the chance constraints \eqref{c:icc} are converted into deterministic constraints \eqref{eq:iccdone}, thereby allowing us to develop the solution algorithm, which we call ITF-CCP, by modifying Algorithm~\ref{alg:itf} as follows:
\begin{itemize}
\item Input: replace $\vec Q$ with $\vec{\hat Q}=\{\hat Q_i\}$ and add $\vec\sigma=\{\sigma_i\}$.
\item Line \ref{alg:init}: replace $ice_i$ and $tank_i$ with $ice_i=c_i \hat Q_i/(\psi_i\Psi)$ and $tank_i=c_i \hat Q_i/(\psi_i\Psi)+(\hat Q_i+\sigma_i z_{\alpha_i})/\psi_i$, respectively.
\end{itemize}

We then run ITF-CCP with $\alpha_i=0.05$ for the same 100 users as in EVM (and also the same realization of $\tilde Q_i$, for a fair comparison).
The new set of results is shown in \fref{fig:ccp}, where we see that there are only 4 cases of over provisioning. This is consistent with the ``exceeding'' probability $\alpha_i$ and shows that the occurrences of service over-provisioning are now {\em under control}.

A side effect is that, as $\hat Q_i'\triangleq\hat Q_i+\sigma_i z_{\alpha_i}<\hat Q_i$, there will be extra resources left when $Q_{ext}\triangleq Q_{tot}-\sumiN\hat Q_i'>0$.
To overcome this, we allow $q_i$ to ``burst'' above $\hat Q_i'$ when $Q_{ext}>0$, but still cap $q_i$ by the actual demand $\tilde Q_i$ in order to avoid over-provisioning. As $\tilde Q_i$ is only realized during service consumption, we allocate $Q_{ext}$ {\em after} a user has consumed his granted quota in the subsequent slot (since $q_i$ is granted at the end of the current slot), based on the first-come-first-serve (FCFS) principle. Note that: (i) incentive is not compromised because obtaining service via a burst is non-guaranteed (opportunistic) as it depends on the availability of $Q_{ext}$ and other users' service consumption, unlike the guaranteed service quota granted by ITF-CCP,
(ii) FCFS does not lead to each user rushing to use up his granted quota in order to take advantage of the burst, because a user will {\em not} know the availability of $Q_{ext}$ until he uses up his granted service quota.

\section{Performance Evaluation}\label{sec:simu}

In this section, we evaluate the performance of the four proposed schemes, IDF, ITF, NE, and ITF-CCP, via simulation. For a more meaningful comparison, we also add two baseline schemes:
\begin{enumerate}
\item Equal Allocation (EA): all the users share the total service quota equally, i.e., $q_i=Q_{tot}/N$.
\item Demand-based Allocation (DA): each user is granted a service quota of $q_i = Q_{tot}\times Q_i/\suml Q_l$ when $\suml Q_l>Q_{tot}$, and $q_i=Q_i$ when $\suml Q_l\le Q_{tot}$.
\end{enumerate}

Similar to \sref{sec:ccp}, the system is hourly-slotted with $N$=100 users. $Q_i=\mathcal U(0,1)$ (hour), $Q_{tot}=\mathcal U(0.5,1)\times\sumiN Q_i$, and the rest of the simulation setup is kept unchanged. In the case of NE, $Q_i^*$ is computed according to \eqref{eq:qne}.
In the case of ITF-CCP, if $Q_{ext}>0$, users burst as follows. Let $t_i\in\mathcal U(q_i,1)$ be the time when a user uses up his granted service quota $q_i$, upon which he can realize via server notification or query through server that there is extra quota, and hence will try to maximize his own benefit by continuously consuming service until he reaches his actual demand $\tilde Q_i$ or $Q_{ext}$ is used up. This is equivalent to running a further round of ITF by setting $Q_{tot}=Q_{ext}$, $ice_i=t_i$, $tank_i=t_i+(\tilde Q_i-t_i)^+$, and $\psi_i=1$.

\subsection{Macro-level Performance}
This sub-section evaluates performance at the system level, in terms of Jain's fairness index as defined by \eref{eq:jrsq}, and social welfare as defined by \eref{eq:swa}.
\begin{figure}[ht]
\centering
\subfloat[Jain's fairness index.]{\label{fig:frsq}
\includegraphics[trim=5mm 2mm 1cm 5mm,clip,width=0.5\linewidth]{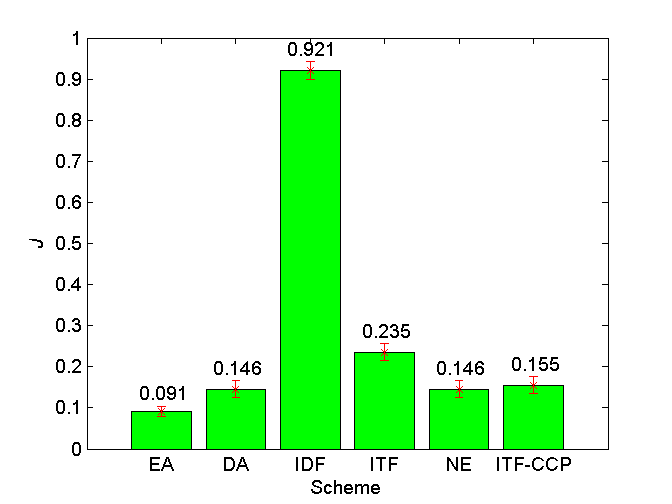}}
\subfloat[Social welfare.]{\label{fig:satis}
\includegraphics[trim=5mm 2mm 1cm 5mm,clip,width=0.5\linewidth]{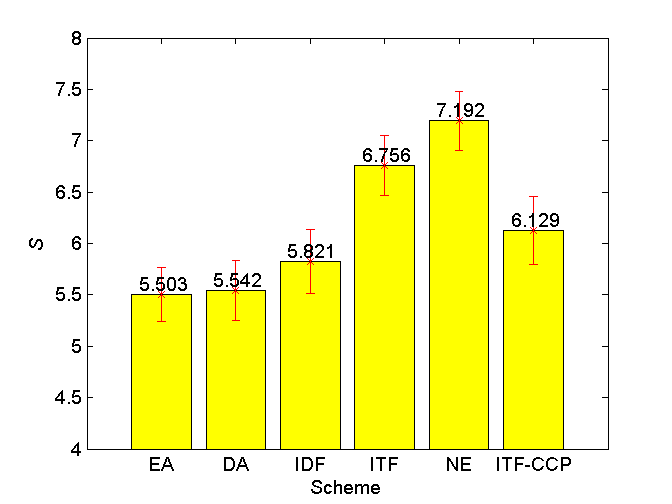}}
\caption{Performance comparison between 6 schemes: EA, DA, IDF, ITF, NE and ITF-CCP. The sample size for each data point is 100. Error-bars represent 95\% confidence intervals.}
\label{fig:comp}
\end{figure}

\fref{fig:comp} presents the results. Each data point is averaged over 100 rounds of simulation, and we also plot upper and lower 95\% confidence limits around the sample means.

In the fairness aspect, \fref{fig:frsq} clearly shows that the IDF scheme outperforms the other schemes and closely approaches the maximum of Jain's index, 1, with a score of 0.92.
As for social welfare (\fref{fig:satis}), the first observation, which is not surprising, is that NE is the clear winner, as is theoretically proven by \thmref{thm:ne}.
On the other hand, the $Q_i^*$ computed by NE may not necessarily reflect users' real needs, and hence other schemes which allow users to declare their demands should still be considered. In that case, ITF is the best scheme and achieves 94\% of the maximum that NE achieves. In the case of coping with uncertainty, ITF-CCP achieves 85.2\% of the NE maximum. The reason for the slight drop is that ITF-CCP takes stricter constraints to avoid over-provisioning when demands are uncertain. Bursting, as an auxiliary mechanism, only helps marginally, because (i) it only takes effect when $Q_{ext}>0$, which is a rare case because $Q_{tot}$ is usually well below $\sumiN Q_i$, (ii) the maximum burst amount for each user is capped by a limited amount of $(\tilde Q_i-t_i)^+$, (iii) FCFS is not optimized for maximizing social welfare (e.g., no priority is given to users with larger marginal utility). These are the trade-offs a service provider should take into consideration when dealing with uncertainty.

Our results also show that none of the schemes is the best in meeting its unintended objective, meaning that there does not exist a ``one-size-fits-all'' solution. Therefore, the correct objective should be carefully considered by a service provider before making a decision on which scheme to employ.

\subsection{Micro-level Performance}

This sub-section zooms in to examine performance at the individual users' level. Specifically, we look at four representative users summarized in \tref{tab:indusers}. The population size is still $N$=100 and the remaining 96 users are all normal users (the same as User 4). As user parameters are fixed in this setting, it makes sense to exclude NE and ITF-CCP from the comparison. System parameters remain the same as in \sref{sec:ccp}.
\begin{table}[ht]
\ifdefined\JNL
  \caption{Unwelcome users vs. a normal user.}\centering
\else
  \caption{Four representative users.}\centering
\fi
  \begin{tabular}{| c | l || l | l | l |} \hline
  User & Type & Demand (hr) & Contrib. (kb) & Cost (\$) \\ \hline\hline
     1 & High-Demand      & \cellcolor[gray]{0.8}1   & 0.5  & 0.5 \\ \hline
     2 & Low-Contribution & 0.5 & \cellcolor[gray]{0.8}0.25 & 0.5 \\ \hline
     3 & High-Cost        & 0.5 & 0.5  & \cellcolor[gray]{0.8}1 \\ \hline
     4 & Normal & 0.5 & 0.5  & 0.5 \\ \hline
  \end{tabular}
\label{tab:indusers}
\end{table}

\begin{figure}[ht]
\centering
\subfloat[Demand-normalized service quota.]{\label{fig:indqQ}
\includegraphics[trim=9mm 6mm 18mm 1cm,clip,width=0.5\linewidth]{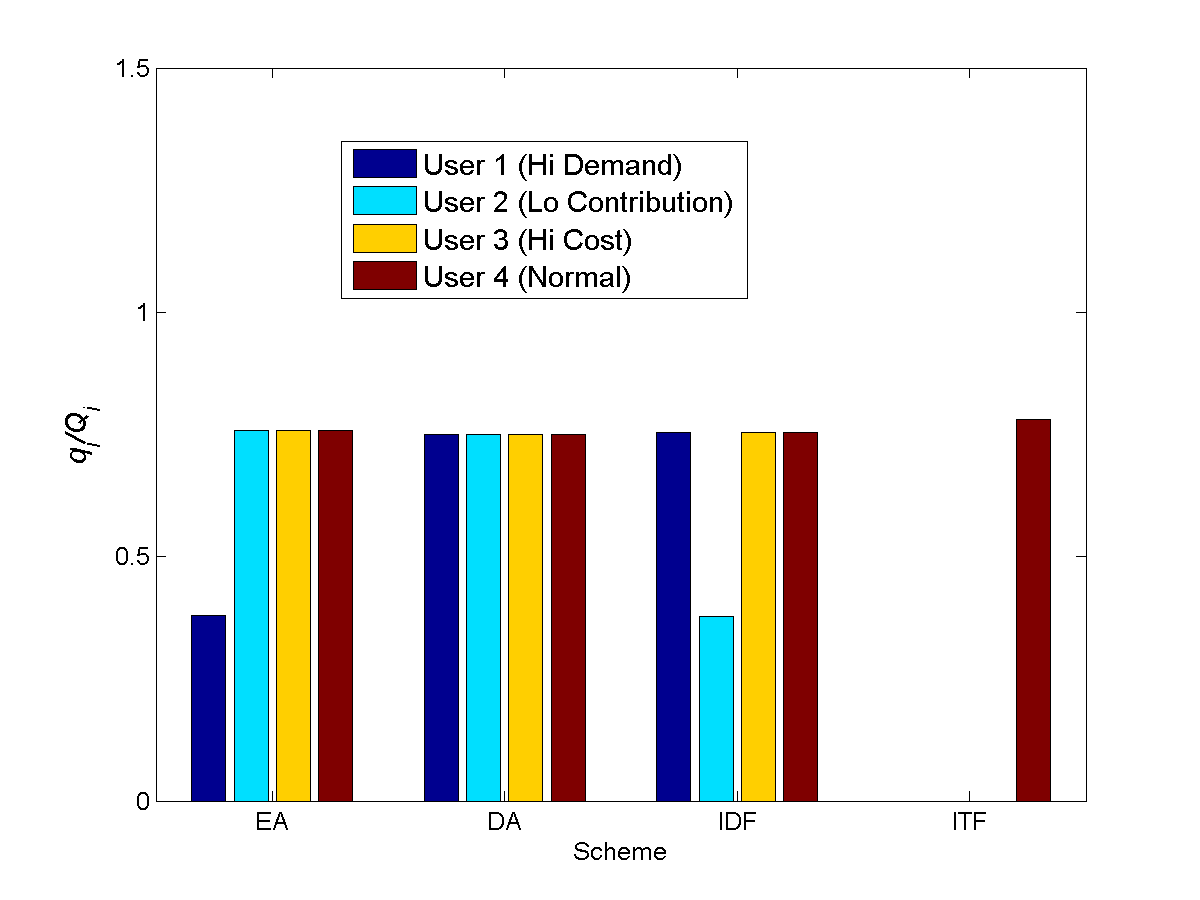}}
\subfloat[Individual user utility.]{\label{fig:indU}
\includegraphics[trim=9mm 6mm 18mm 1cm,clip,width=0.5\linewidth]{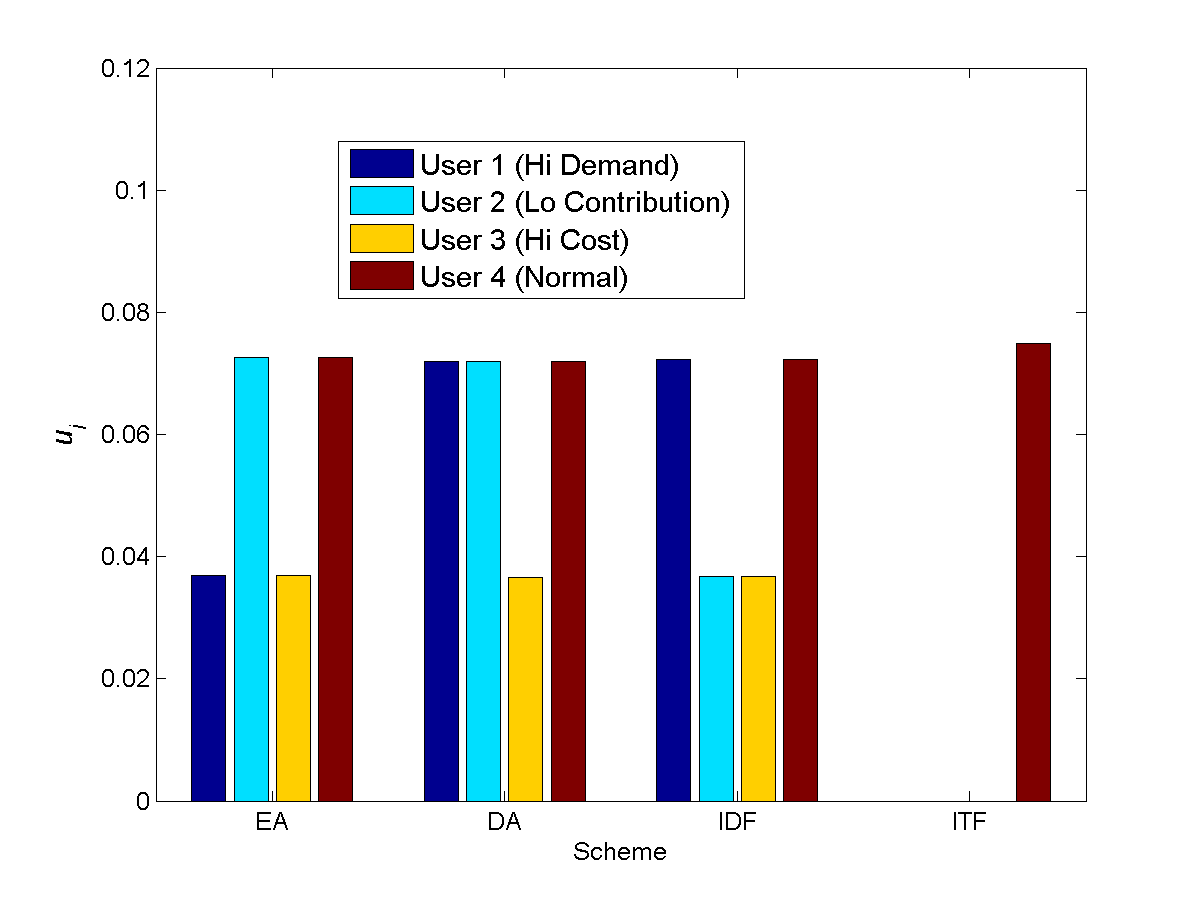}}
\ifdefined\JNL
	\caption{Performance of unwelcome users vs. normal users.}
\else
	\caption{Performance of four representative users.}
\fi
\label{fig:ind1}
\end{figure}

In \fref{fig:indqQ}, we compare the service quota each user received against his demand, i.e., $q_i/Q_i$. Under IDF, all users except for User 2 reaches $\sim$75\% which is the total service availability level (recall that $Q_{tot}=0.75\sumiN Q_i$). User 2 contributes only half of what the other users contributed, and, in return, he is rewarded by 37.5\%, which is also half that of the others. This indicates an incentive that encourages higher user contributions. User 1 benefits more in terms of {\em absolute} service quota, $q_i$, because IDF does not discriminate against high-demand users. In sharp contrast, ITF grants users 1--3 zero service. The reason is that they are classified as ``hard-to-satisfy'' or ``unwelcome'' users, following the philosophy of ITF. As such, priority is given to the remaining 97 (normal) users who equally share $Q_{tot}$ and obtain a $q_i/Q_i$ even slightly higher than 75\%.

\fref{fig:indU} gives each user's utility. From the above, it is easy to understand that, under ITF, only User 4 receives positive utility which is also slightly higher than the maximum of other schemes. Under IDF, User 2 receives lower utility corresponding to the lower $q_i$ as in \fref{fig:indqQ}, User 3 receives lower utility because of his higher cost as per the definition of utility in \eref{eq:swa},\footnote{The approximately linear relationship in spite of the existence of $\log(\cdot)$ is because $\Psi\frac{q_i}{c_i Q_i}$ is small and $\log(1+x)\sim x$ for small $x$.} User 1 receives the same utility as normal users because IDF tries to make $q_i/Q_i$ commensurate with $\psi_i$ and thereby removes the difference between User 1 and normal users.

It is worth noting that, while User 2 is under-privileged in both IDF and ITF, he (undesirably) obtains the same amount of service as normal users under EA and DA. This clearly demonstrates the advantage of the built-in incentive mechanism in our designed schemes.

\ifdefined\JNL

Now, we investigate the other scenario with ``welcome'' users to see if and how they benefit from the schemes that we propose. \tref{tab:indusers2} gives the user setting and the remaining users (5--100) are the same as the normal user (User 4).
\begin{table}[ht]
  \caption{Welcome users vs. a normal user.}\centering
  \begin{tabular}{| c | l || l | l | l |} \hline
  User & Type & Demand (hr) & Contrib. (kb) & Cost (\$) \\ \hline\hline
     1 & Low-Demand      & \cellcolor[gray]{0.8}0.25   & 0.5  & 0.5 \\ \hline
     2 & High-Contribution & 0.5 & \cellcolor[gray]{0.8}1 & 0.5 \\ \hline
     3 & Low-Cost        & 0.5 & 0.5  & \cellcolor[gray]{0.8}0.25 \\ \hline
     4 & Normal & 0.5 & 0.5  & 0.5 \\ \hline
  \end{tabular}
\label{tab:indusers2}
\end{table}

The simulation results are given in \fref{fig:ind2}. Regarding demand-normalized service quota, we see that User 1 achieves $q_i/Q_i\approx 1.5$ under the EA scheme, which apparently results in resource waste. On the other hand, DA equalizes $q_i/Q_i$ for all users, thereby not motivating any ``good'' users. IDF rewards User 2, as desired, while not User 1 and 3, because it aims to fairly allocate $q_i/Q_i$ only and does not take cost into account. ITF, at last, can be viewed as the best scheme among the four because all the good users (1--3) receive higher demand-normalized service quota than normal users, which serves as a good incentive.

\begin{figure}[ht]
\centering
\subfloat[Service quota relative to user demand.]{\label{fig:indqQ2}
\includegraphics[trim=9mm 6mm 18mm 1cm,clip,width=0.7\linewidth]{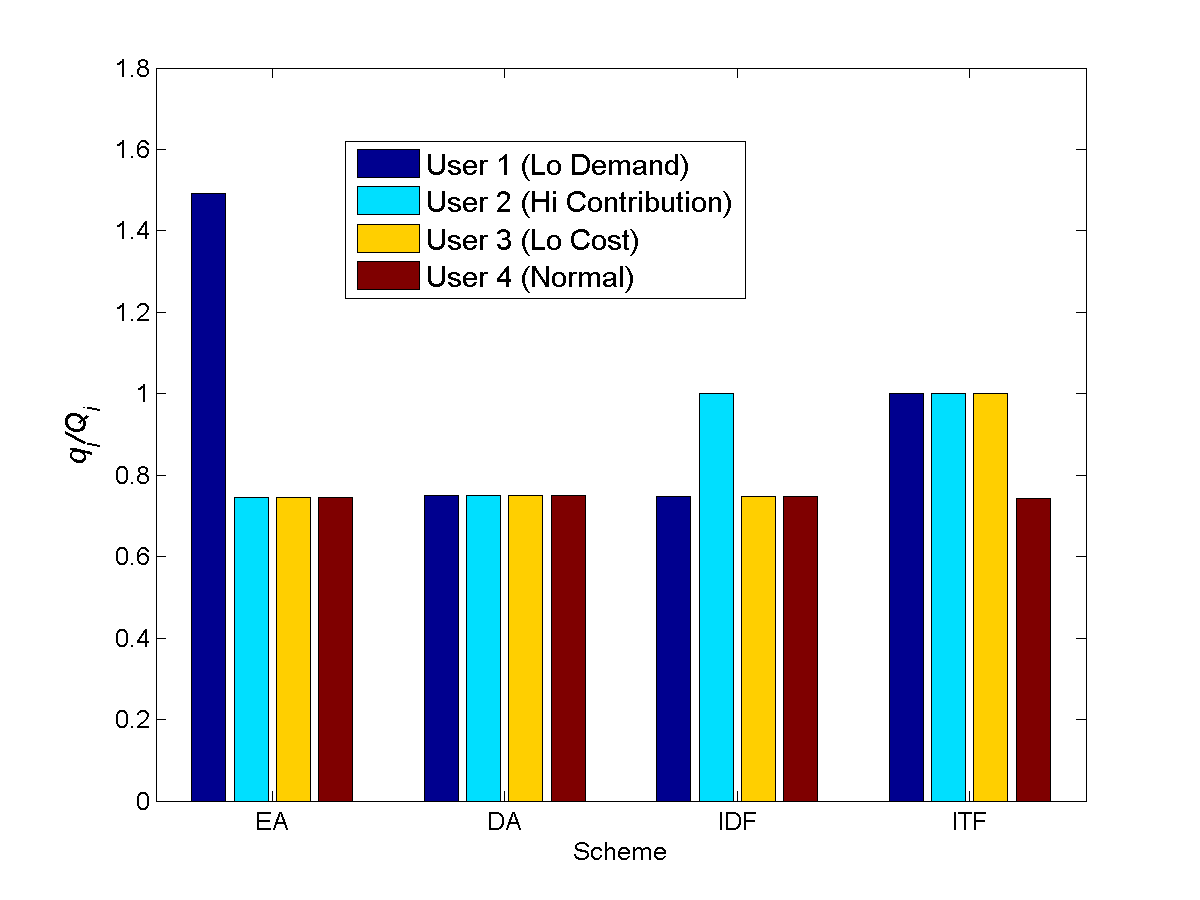}}\vfil
\subfloat[Individual user utility.]{\label{fig:indU2}
\includegraphics[trim=9mm 6mm 18mm 1cm,clip,width=0.7\linewidth]{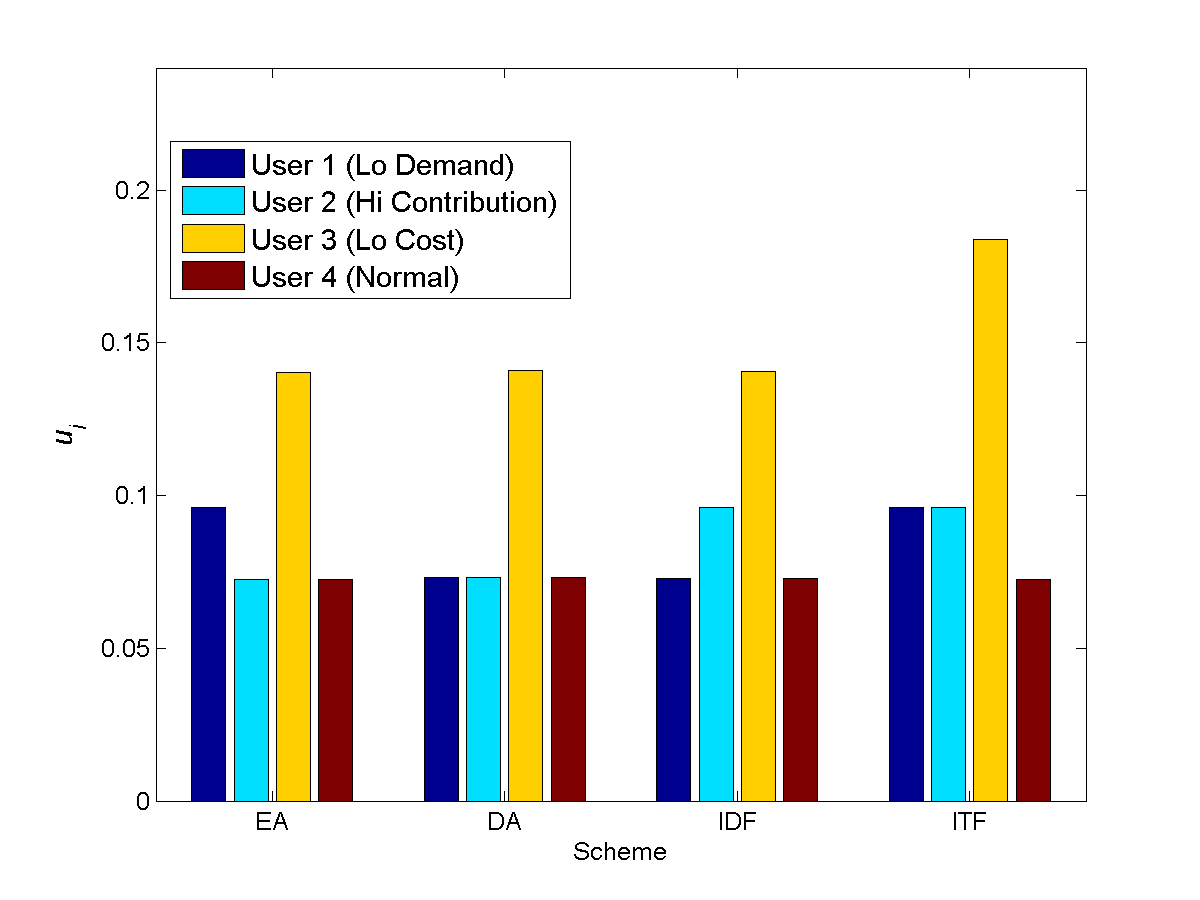}}
\caption{Performance of welcome users vs. normal users.}
\label{fig:ind2}
\end{figure}

With respect to user utility (\fref{fig:indU2}), we see that User 2, though contributing highly, is not better off than normal users under EA and EA. This will, adversely, dampen the morale of good contributors. On the contrary, IDF and ITF both show affinity to User 2 and, among these two schemes, ITF also satisfies User 1 and 3 more than IDF, which is desired and can be understood by comparing with \fref{fig:indqQ2}. The reason that User 3 receives apparently higher utility than other users is because of the definition of utility (\eref{eq:utdef}) and the values of $q_i/Q_i$ shown in \fref{fig:indqQ2}.

Finally, we note that the approximately linear relationship between \fref{fig:indqQ} and \fref{fig:indU} is not reproduced between \fref{fig:indqQ2} and \fref{fig:indU2}. This is because $\Psi\frac{q_i}{c_i Q_i}$ is no longer small enough in this case, which renders the rule of $\log(1+x)\sim x$ inapplicable.

\fi

\section{Conclusion}\label{sec:conc}

We address the issue of incentive in participatory sensing as it is of pivotal importance for actualizing this new sensing paradigm. Instead of using monetary or reputational incentives, we take a demand-based approach and motivate users through regulated quantities of compelling services that they desire to consume. To this end, we designed two schemes, IDF and ITF, to address the key questions of how to incentivize users to contribute and, at the same time, maximize fairness and social welfare. Our theoretical and simulation investigations both demonstrated the effectiveness of our designed schemes and their desired properties. Two tailored variations, NE and ITF-CCP, were also presented for two other scenarios of interest, using a game-theoretic approach and a stochastic programming technique, respectively.

\bibliographystyle{IEEEtran}
\bibliography{IEEEabrv,netecon}

\end{document}